\documentclass{article}

\usepackage{PRIMEarxiv}

\usepackage[utf8]{inputenc} % allow utf-8 input
\usepackage[T1]{fontenc}    % use 8-bit T1 fonts
\usepackage{hyperref}       % hyperlinks
\usepackage{url}            % simple URL typesetting
\usepackage{booktabs}       % professional-quality tables
\usepackage{amsfonts}       % blackboard math symbols
\usepackage{nicefrac}       % compact symbols for 1/2, etc.
\usepackage{microtype}      % microtypography
\usepackage{lipsum}
\usepackage{fancyhdr}       % header
\usepackage{graphicx}       % graphics
\usepackage{caption}
\usepackage{tabulary} 
\usepackage[flushleft]{threeparttable}
\usepackage{multirow}
\usepackage{adjustbox}
\graphicspath{{media/}}     % organize your images and other figures under media/ folder
\usepackage{subcaption}
\usepackage{amsmath,amsfonts,bm}
\usepackage{lineno}
\usepackage[toc,page]{appendix}
\usepackage[colorinlistoftodos,prependcaption,textsize=tiny]{todonotes}
\usepackage{float}  % for the [H] option
\usepackage{algorithm} 
\usepackage{algpseudocode} 
\usepackage{makecell}
\usepackage[normalem]{ulem}
\usepackage{hyperref}
\usepackage{subcaption}
\usepackage{titling}

\title{Technical Report of HelixFold3 \\for Biomolecular Structure Prediction}

\author{
Lihang Liu$^1$, Shanzhuo Zhang$^1$, Yang Xue$^1$, Xianbin Ye$^1$, Kunrui Zhu$^1$, Yuxin Li$^1$, Yang Liu$^1$, Jie Gao$^1$, \\
Wenlai Zhao$^{2,3}$, Hongkun Yu$^{2,3}$, Zhihua Wu$^2$, Xiaonan Zhang$^1$, Xiaomin Fang$^1$\thanks{Corresponding author: Xiaomin Fang (fangxiaomin01@baidu.com)} \\
$^1$PaddleHelix Team, Baidu Inc., $^2$Tecorigin Ltd, $^3$Tsinghua Unv \\
  Official website: \texttt{\url{https://paddlehelix.baidu.com/}} 
}
\date{December 19, 2024}

\begin{document}
\maketitle
%\linenumbers

\begin{abstract}
The AlphaFold series has transformed protein structure prediction with remarkable accuracy, often matching experimental methods. AlphaFold2, AlphaFold-Multimer, and the latest AlphaFold3 represent significant strides in predicting single protein chains, protein complexes, and biomolecular structures. While AlphaFold2 and AlphaFold-Multimer are open-sourced, facilitating rapid and reliable predictions, AlphaFold3 remains partially accessible through a limited online server and has not been open-sourced, restricting further development.

To address these challenges, the PaddleHelix team is developing HelixFold3, aiming to replicate AlphaFold3's capabilities. Leveraging insights from previous models and extensive datasets, HelixFold3 achieves accuracy comparable to AlphaFold3 in predicting the structures of the conventional ligands, nucleic acids, and proteins. The initial release of HelixFold3 is available as open source on GitHub for academic research, promising to advance biomolecular research and accelerate discoveries. The latest version will be continuously updated on the HelixFold3 web server, providing both interactive visualization and API access.

\end{abstract}

\section{Introduction}
AlphaFold series \cite{jumper2021highly, evans2021protein,abramson2024accurate} revolutionizes protein structure prediction with unprecedented accuracy, often rivaling experimental methods. AlphaFold2 \cite{jumper2021highly}, AlphaFold-Multimer \cite{evans2021protein}, and AlphaFold3 \cite{abramson2024accurate}, achieve breakthrough progress in the prediction of single protein chains, protein complexes, and biomolecular structures. Both AlphaFold2 and AlphaFold-Multimer have fully open-sourced their codes, significantly accelerating protein-related research by providing rapid and reliable predictions. These tools not only enhance our understanding of protein functions and interactions, but also exemplify the transformative potential of artificial intelligence in solving complex scientific challenges.

AlphaFold3, the latest in the series, supports biomolecular interaction predictions and offers an online server \footnote{https://alphafoldserver.com/} for limited structural prediction services. This server allows researchers to utilize its advanced capabilities, although it cannot support arbitrary biomolecular structure predictions and imposes a daily limit on the number of predictions. Furthermore, AlphaFold3 has not yet been open-sourced, limiting its accessibility for widespread use and further development by the scientific community.

Replicating AlphaFold3's capabilities presents significant opportunities for advancing the life sciences but involves substantial challenges due to the model's complexity, data requirements, and the extensive computational resources needed for training.

The PaddleHelix team is working on HelixFold3 with the objective of replicating the advanced capabilities of AlphaFold3. Our approach is informed by insights from the AlphaFold3 paper and builds on our prior work with HelixFold \cite{wang2022helixfold}, HelixFold-Single \cite{fang2023method}, HelixFold-Multimer \cite{fang2024helixfold}, and HelixDock \cite{liu2023pre}. For training, we utilized targets from the Protein Data Bank (PDB) \cite{berman2000proteindatabank} released before September 30, 2021, along with self-distillation datasets. Currently, HelixFold3's accuracy in predicting the structures of small molecule ligands, nucleic acids (including DNA and RNA), and proteins is comparable to that of AlphaFold3. We are committed to continuously enhancing the model's performance and rigorously evaluating it across a broader range of biological molecules.

The initial release of HelixFold3, which includes the inference code and the current version of model parameters, is now available as open source on PaddleHelix's GitHub repository. You can access it at \texttt{\href{https://github.com/PaddlePaddle/PaddleHelix/blob/dev/apps/protein_folding/helixfold3}{https://github.com/PaddlePaddle/PaddleHelix/blob/dev/apps/protein\_folding/helixfold3}} for academic research. This release is intended for non-commercial use and provides researchers with the tools needed to explore and leverage HelixFold3's advanced capabilities in biomolecular structure prediction. We believe that the open-source availability of HelixFold3 will significantly contribute to the advancement of research in biomolecular interactions. We also provide online service at PaddleHelix website at \texttt{\href{https://paddlehelix.baidu.com/app/all/helixfold3/forecast}{https://paddlehelix.baidu.com/app/all/helixfold3/forecast}}, providing both interactive visualization and API access. HelixFold3 also provides a simple and user-friendly API service (\texttt{\href{ https://paddlehelix.baidu.com/app/tut/guide/all/helixfold3sdk}}). Users can easily schedule Baidu Intelligent Cloud resources with just a few simple steps, enabling rapid execution of tens of thousands of structure predictions without the need for expensive hardware or complex configurations. The API supports seamless integration with existing bioinformatics toolchains and is widely applicable across various fields of scientific research and commercial applications.

%Researchers can now build upon this foundation, conduct further studies, and apply HelixFold3 to a broader range of biological questions, thereby advancing our understanding of complex biomolecular systems and accelerating the development of new applications in structural biology and related areas.

\section{Results}
Due to ongoing iterations of HelixFold3, we have labeled the version released on GitHub in August 2024 as HelixFold3, and the version released in December 2024 on the PaddleHelix web server as HelixFold3.1 in subsequent results.

We begin by evaluating the performance of HelixFold3 in multiple datasets. For ligands, we utilize the PoseBusters benchmark \cite{buttenschoen2024posebusters} to assess precision and physical plausibility. For nucleic acids, HelixFold3 is evaluated on CASP15 \cite{kryshtafovych2023critical} RNA targets and recent RNA, DNA, and protein-RNA complex structures from the RCSB Protein Data Bank (RCSB PDB) \cite{berman2000proteindatabank}. We also evaluated the model's precision in predicting protein structures using recently released protein-protein complexes from the PDB and the SAbDab database \cite{James2014sabdab}, which serves as an antigen-antibody evaluation set. Each sample is processed with 5 different random seeds, and diffusion inference is performed 5 times per seed with 200 sampling steps. The prediction with the highest confidence score was used for the evaluation.

In addition, we investigate the effectiveness of confidence metrics in evaluating prediction quality. This involves analyzing how well the confidence scores correlate with the actual accuracy of predictions across various datasets. We also explore how different factors impact prediction quality, including the number of random seeds, diffusion inference iterations, and the number of sampling steps in the diffusion process. This comprehensive analysis aims to refine our understanding of how these parameters influence the reliability of the prediction.

\subsection{Ligands}

\begin{figure}[h]
    \centering
    \begin{subfigure}[b]{\textwidth}
        \centering
        \includegraphics[width=0.95\textwidth]{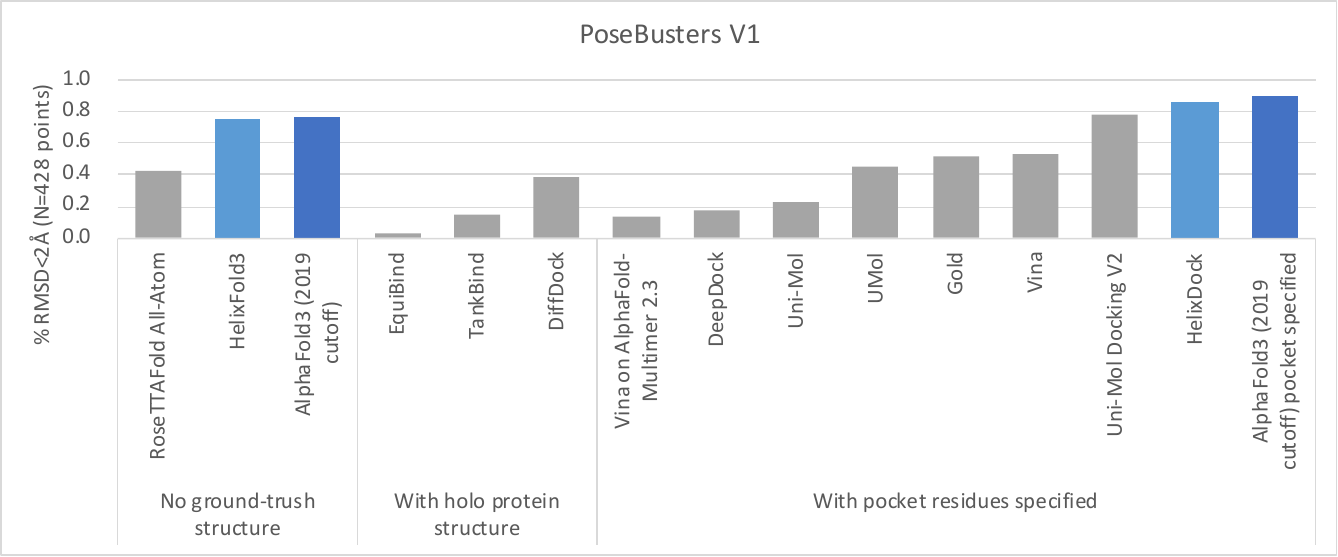}
        \caption{Comparison of the success rate of HelixFold3 and baselines on the PoseBusters(V1) with 428 targets.}
        \label{fig:ligands_posebusters_v1}
    \end{subfigure}
    \hfill
    \begin{subfigure}[b]{0.95\textwidth}
        \centering
        \includegraphics[width=\textwidth]{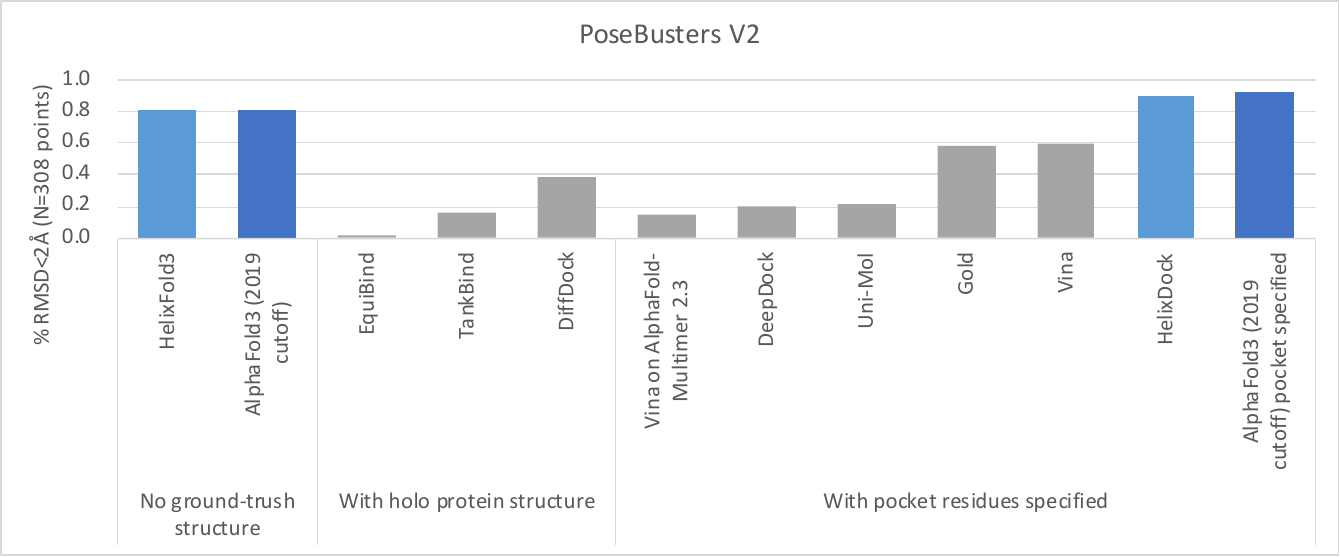}
        \caption{Comparison of the success rate of HelixFold3 and baselines on the PoseBusters(V2) with 308 targets.}
        \label{fig:ligands_posebusters_v2}
    \end{subfigure}
    \begin{subfigure}[b]{0.95\textwidth}
        \centering
        \includegraphics[width=\textwidth]{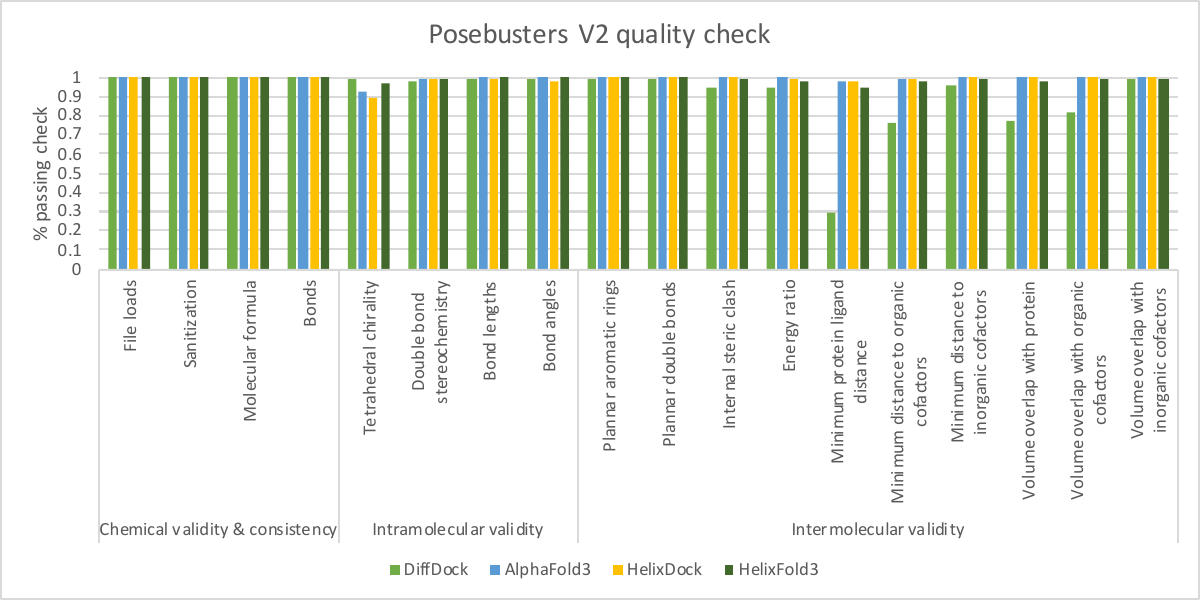}
        \caption{PoseBusters Quality check.}
        \label{fig:ligands_quality_check}
    \end{subfigure}
    \caption{Results for ligands.}
    \label{fig:ligands}
\end{figure}

We first compare HelixFold3 and the baseline methods in PoseBusters \cite{buttenschoen2024posebusters} to evaluate the quality of the protein-ligand structure predictions. The PoseBusters dataset, a benchmark for ligand docking algorithms, initially had 428 structures (PoseBusters V1). After excluding data points with ligands within 5.0Å of multiple biological units, it was refined to 308 structures (PoseBusters V2). The baseline methods can be classified into three groups: methods with no ground-truth protein structure specific, methods with holo protein structure specified, and methods with pocket residues specified. The comparison of the success rate for PoseBuseters V1 and PoseBuseters V2 is shown in Figure~\ref{fig:ligands_posebusters_v1} and Figure~\ref{fig:ligands_posebusters_v2}. Even though no ground-truth protein structure is specified, HelixFold3 achieves a high success rate, surpassing methods that rely on given homo protein structures. Its prediction accuracy is comparable to that of AlphaFold3, highlighting its strong performance in predicting protein-ligand interactions. Due to the overlap between the training data for HelixFold3 and PoseBusters, there is a risk of overestimating HelixFold3's performance. To address this, we evaluated the average success rate on 290 samples that do not overlap with the training data. This analysis showed only a 2\% reduction in performance, suggesting that HelixFold3 maintains strong accuracy in predicting ligand-protein interface structures.

%To assess HelixFold3’s precision across varying protein target difficulties, we stratified the PoseBusters V2 set based on the target protein receptor’s maximum sequence identity with proteins in the PDBbind 2020 General Set \cite{wang2004pdbbind}, using the methodology outlined in \cite{buttenschoen2024posebusters}. The test cases were categorized into three groups by sequence identity: low [0\%, 30\%], medium (30\%, 90\%], and high (90\%, 100\%]. As depicted in Figure~\ref{fig:ligands_sequence_identity}, physics-based methods, maintain consistent performance across all three groups. In contrast, most deep-learning-based methods, experience a performance decline with proteins of lower sequence identity. HelixFold3 also experiences a slight decline in success rate as sequence identity decreases. However, HelixDock, which was pre-trained on large-scale protein-ligand data, exhibits consistent performance. To enhance HelixFold3’s robustness, we plan to incorporate protein-ligand structural data in future updates.

To evaluate the stereochemistry and physical plausibility of the predicted ligand structures, including intra- and intermolecular measurements, we used the PoseBusters test suite \cite{buttenschoen2024posebusters}. As shown in Figure~\ref{fig:ligands_quality_check}, HelixFold3, AlphaFold3, and HelixDock all achieve pass rates above 90\% for nearly all metrics, with the exception of tetrahedral chirality. %We illustrate with a case (Figure~\ref{fig:ligands_case}) that increasing the sampling steps of the diffusion module enhances the validity of the predicted structures. Furthermore, increasing the number of random seeds also improves the reasonableness of the predicted structures, which will be discussed in more detail in the following subsections.

\subsection{Nucleic Acids}
\begin{figure}[h]
    \centering
    \begin{subfigure}[b]{0.67\textwidth}
        \centering
        \includegraphics[width=\textwidth]{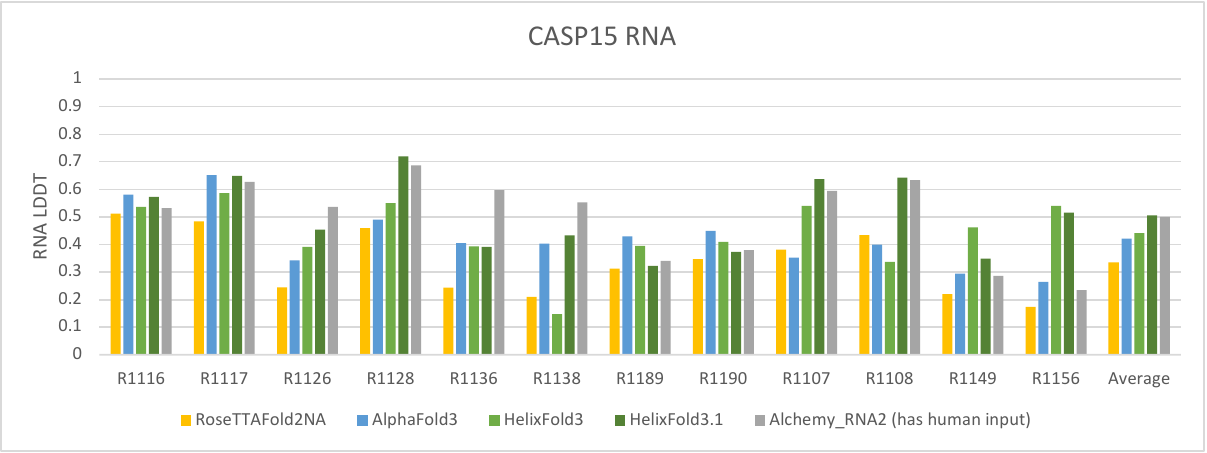}
        \caption{Comparison of RNA LDDT on the 12 RNA targets in CASP15 with available ground-truth structures in PDB.}
        \label{fig:na_casp15}
    \end{subfigure}
    \hfill
    \begin{subfigure}[b]{0.315\textwidth}
        \centering
        \includegraphics[width=\textwidth]{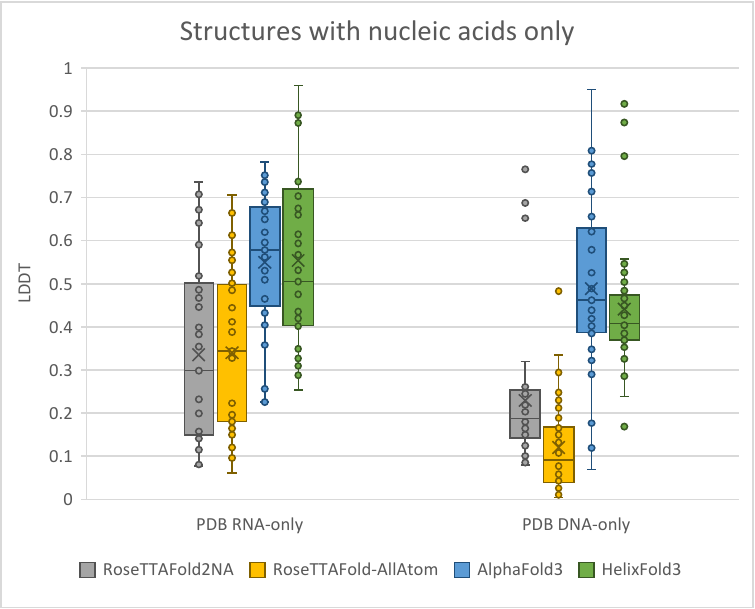}
        \caption{Comparison on RNA-only or DNA-only complexes from PDB.}
        \label{fig:na_recent}
    \end{subfigure}
    \begin{subfigure}[b]{0.5\textwidth}
        \centering
        \includegraphics[width=\textwidth]{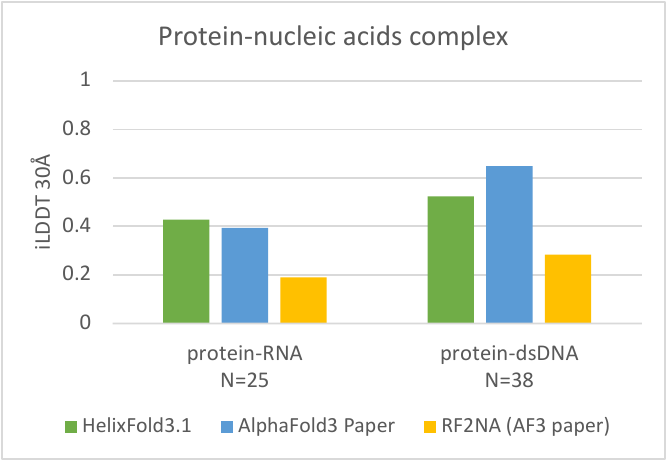}
        \caption{Result for protein-nucleic acid complex.}
        \label{fig:prot_na}
    \end{subfigure}
    \caption{Results for nucleic acid targets.}
    \label{fig:na}
\end{figure}

Accurately predicting the structure of nucleic acid targets in a fully automated manner, without human input, remains a formidable challenge, primarily due to the limited availability of crystallized nucleic acid structures. To evaluate HelixFold3's capabilities in this domain, we conducted a comparative analysis against several baseline methods, starting with RNA targets from CASP15 \cite{kryshtafovych2023critical}, using the evaluation framework established by AlphaFold3. Figure~\ref{fig:na_casp15} illustrates a comparison between HelixFold3 and the baseline methods, showcasing the average RNA LDDT across 12 targets as well as the RNA LDDT for each individual target. HelixFold3.1 outperforms other fully automated models and stands on par with AIchemy\_RNA2 \cite{chen2023rna}, which benefits from human intervention.
% While HelixFold3's accuracy on CASP15 RNA samples does not yet match that of AIchemy\_RNA2 \cite{chen2023rna}, which benefits from human intervention, it achieves a level of accuracy on par with AlphaFold3 among models that operate in a fully automated manner.

Given the limited number of RNA targets in CASP15, we further expanded our evaluation by collecting 41 RNA-only and 41 DNA-only complexes released between May 1, 2022, and June 30, 2024, from the Protein Data Bank (PDB) to more thoroughly assess HelixFold3's performance in nucleic acid structure prediction. The results, depicted in Figure~\ref{fig:na_recent}, demonstrate that HelixFold3 significantly outperforms RoseTTAFold2NA \cite{baek2024accurate}, a model specifically designed for nucleic acid target structure prediction, as well as RoseTTAFold-AllAtom, an all-atom biomolecular structure prediction model.

We also evaluate the accuracy of HelixFold3 on the protein-nucleic acid complexes, using the iLDDT metric within a 30Å range for local structures. As illustrated in Figure \ref{fig:prot_na}, HelixFold3 performs comparable to AlphaFold3 and significantly outperforms RF2NA in predicting protein-RNA and protein-double-stranded DNA complexes. Furthermore, HelixFold3 slightly outperforms AlphaFold3 in protein-RNA complexes.Note that the results for AlphaFold3 are taken directly from the AlphaFold3 paper, making the comparison not entirely fair.

\subsection{Proteins}
\begin{figure}[ht]
    \centering
    \begin{subfigure}[b]{0.4\textwidth}
        \centering
        \includegraphics[width=\textwidth]{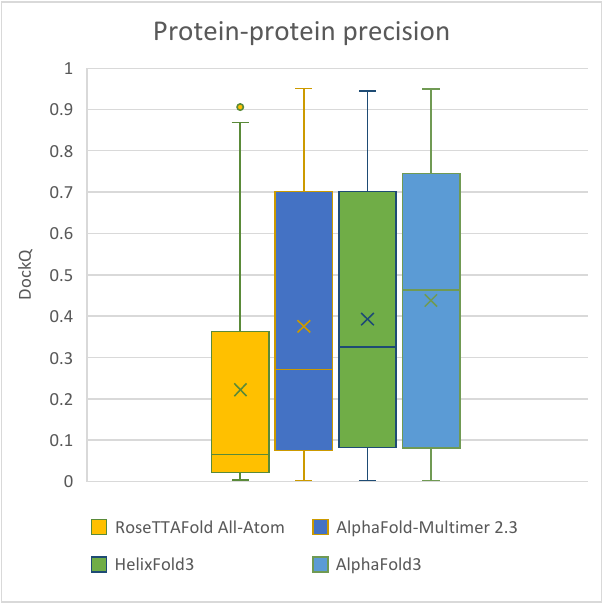}
        \caption{DockQ of protein-protein interface targets collected from PDB.}
        \label{fig:proteins_heter_v2_dockq}
    \end{subfigure}
    %\hfil
    \quad
    \begin{subfigure}[b]{0.4\textwidth}
        \centering
        \includegraphics[width=\textwidth]{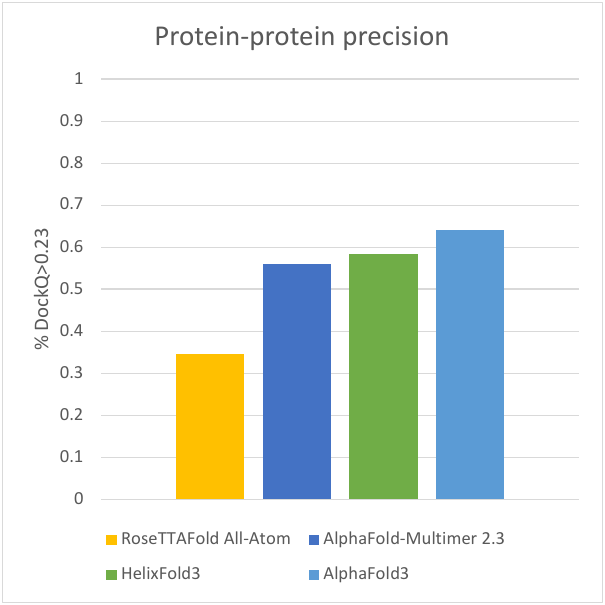}
        \caption{Success rate of protein-protein interface targets collected from PDB.}
        \label{fig:proteins_heter_v2_success_rate}
    \end{subfigure}
    \begin{subfigure}[b]{0.56\textwidth}
        \centering
        \includegraphics[width=\textwidth]{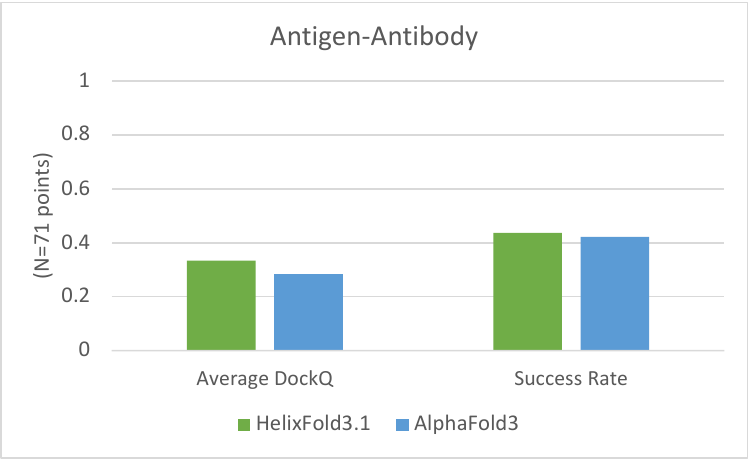}
        \caption{Prediction precise of antigen-antibody collected from SAbDab\cite{James2014sabdab}.}
        \label{fig:abag_base}
    \end{subfigure}
    %\hfil
    \quad
    \begin{subfigure}[b]{0.4\textwidth}
        \centering
        \includegraphics[width=\textwidth]{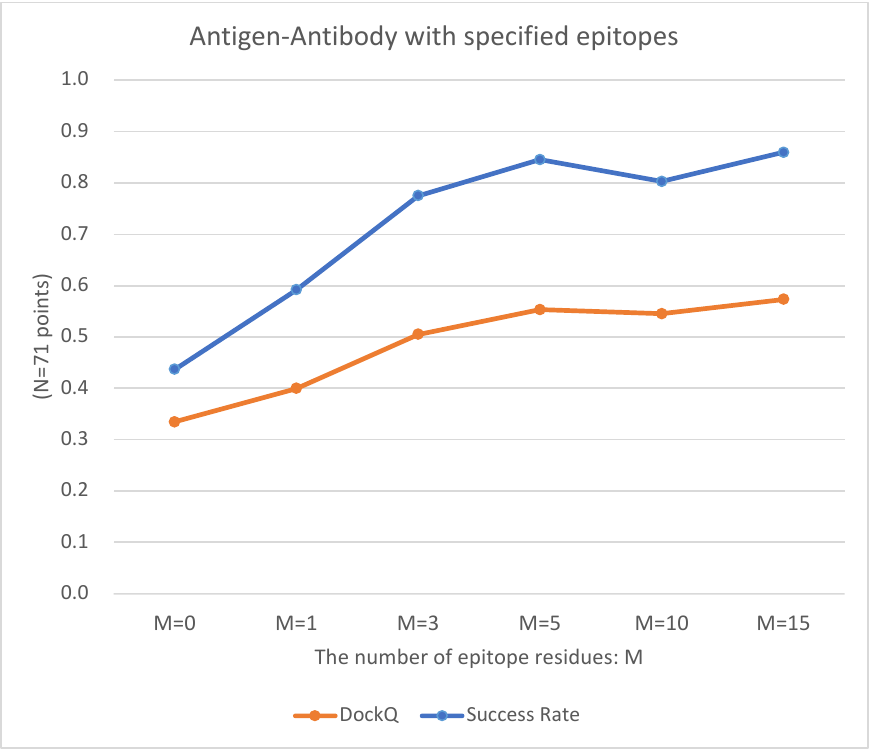}
        \caption{Prediction precise of antigen-antibody with different numbers of specified epitope residues}
        \label{fig:abag_eptiope}
    \end{subfigure}
    \caption{Results for protein targets.}
    \label{fig:proteins}
\end{figure}

For protein-protein complex structure prediction, AlphaFold-Multimer represents a significant advancement over earlier models, though its success rate and accuracy still have room for improvement. AlphaFold3 further enhances these capabilities, delivering superior predictive performance. We analyzed 186 protein complexes released in the PDB from January 19, 2022, to November 30, 2022, to assess HelixFold3 and the competitive methods.  As depicted in Figure~\ref{fig:proteins}, HelixFold3 has already outperformed AlphaFold-Multimer in predicting protein-protein interfaces, yet a gap remains between HelixFold3 and AlphaFold3. To address this, ongoing research will concentrate on targeted optimizations and iterative refinements of HelixFold3, with the aim of achieving greater accuracy and reliability in protein-protein complex predictions.

Antigen-antibody prediction remains a significant challenge in protein complex structure prediction. To evaluate performance in this domain, samples were selected from the SAbDab database \cite{James2014sabdab}, spanning release dates from January 25, 2023, to August 9, 2023.
As shown in Figure \ref{fig:abag_base}, HelixFold3 surpasses AlphaFold3 in two key metrics: DockQ and Success Rate (the percentage of cases with DockQ $\geq 0.23$). By leveraging a small number of specified epitope residues, HelixFold3 achieves higher success rates, as illustrated in Figure \ref{fig:abag_eptiope}. Notably, with the specification of only five epitope residues, HelixFold3 attains a success rate exceeding 80\%.

\subsection{Covalent Modification}
\begin{figure}[ht]
    \centering
    \includegraphics[width=0.8\textwidth]{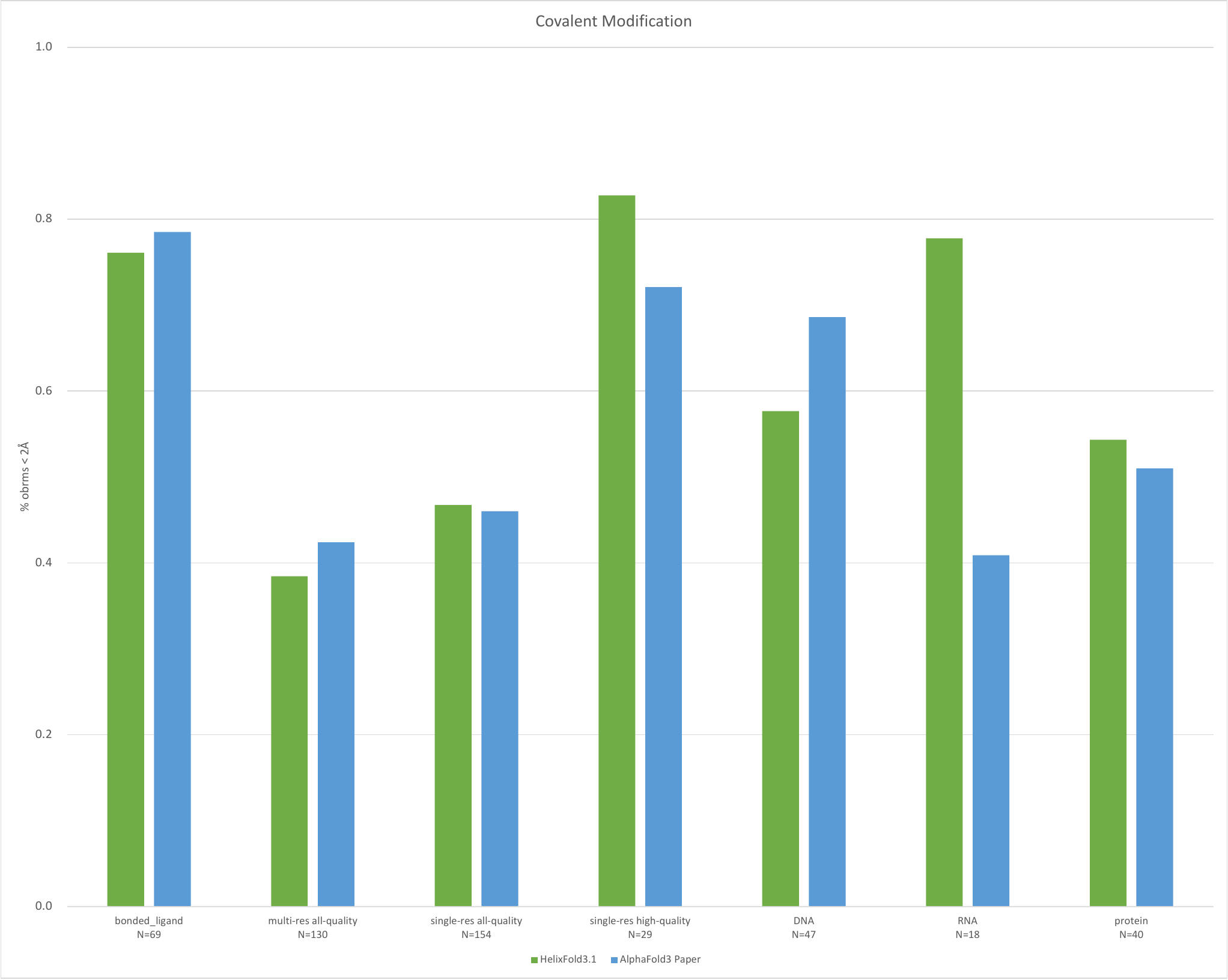}
    \caption{Results for covalent modification}
    \label{fig:coval_modif}
\end{figure}

Covalent modifications are essential for the regulation of protein function, stability, and interactions, yet predicting their structural impact remains a significant challenge for computational methods. Following the evaluation protocol outlined in the AlphaFold3 paper, MMseqs was used to cluster covalently modified structures sourced from RecentPDB. It should be noted that our cluster counts differ slightly from those reported in the AlphaFold3 paper, probably due to minor differences in implementation details or database versions. As a result, direct comparisons of performance may not be entirely fair.
As shown in Figure \ref{fig:coval_modif}, HelixFold3 demonstrates a higher precision than AlphaFold3 in predicting high quality single-residue glycosylation and RNA modification structures, measured by the percentage of successful predictions with RMSD <2Å. For other modifications, both models exhibit comparable performance.

\subsection{Model Confidence}
\begin{figure}[ht]
    \centering
    \begin{subfigure}[b]{0.4\textwidth}
        \centering
        \includegraphics[width=\textwidth]{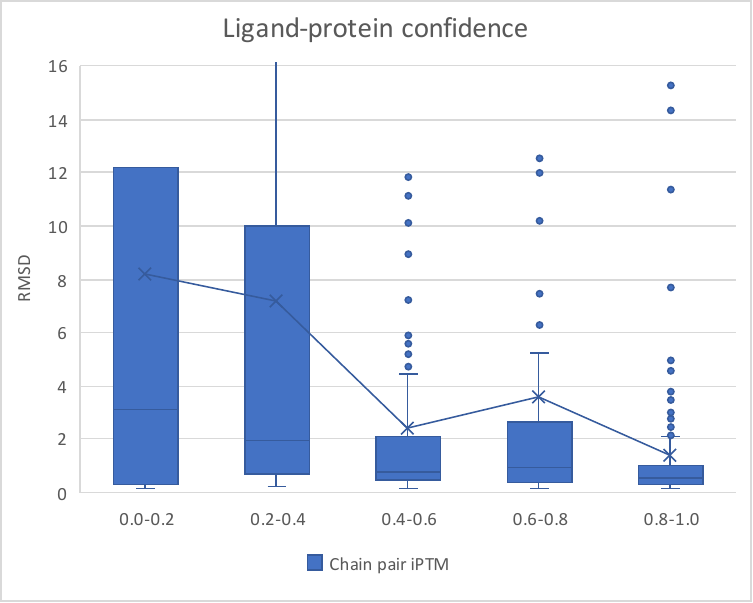}
        \caption{Relation of iPTM and RMSD for ligand-protein interfaces.}
        \label{fig:confidence_ligand_protein}
    \end{subfigure}
    %\hfill
    \quad
    \begin{subfigure}[b]{0.4\textwidth}
        \centering
        \includegraphics[width=\textwidth]{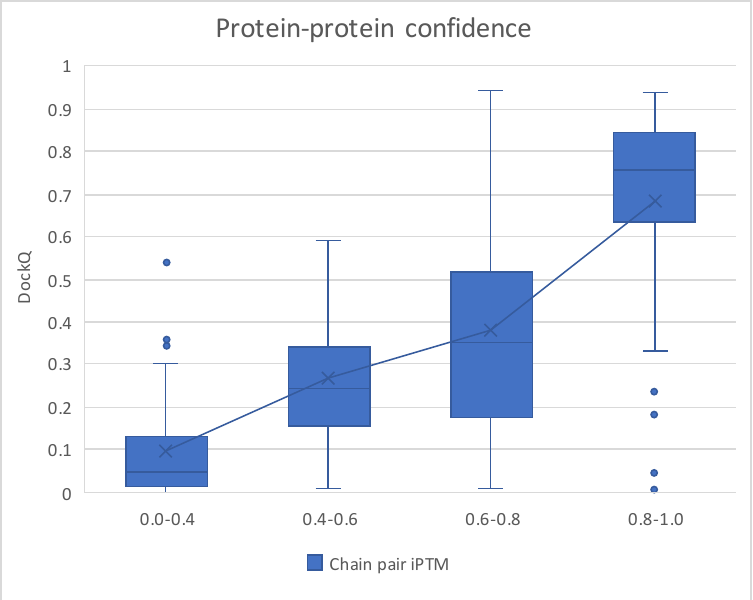}
        \caption{Relation of iPTM and DockQ for protein-protein interfaces.}
        \label{fig:confidence_protein_protein}
    \end{subfigure}
    \begin{subfigure}[b]{0.4\textwidth}
        \centering
        \includegraphics[width=\textwidth]{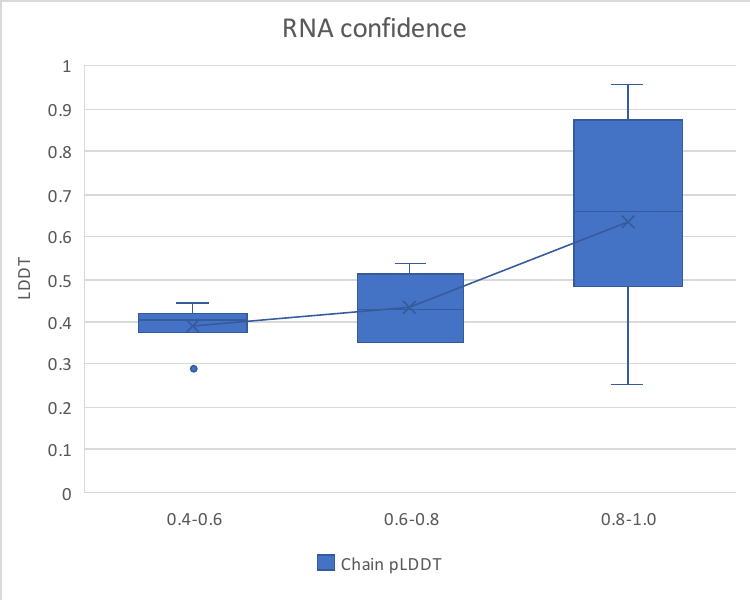}
        \caption{Relation between pLDDT and LDDT for RNA targets.}
        \label{fig:confidence_rna}
    \end{subfigure}
    %\hfill
    \quad
    \begin{subfigure}[b]{0.4\textwidth}
        \centering
        \includegraphics[width=\textwidth]{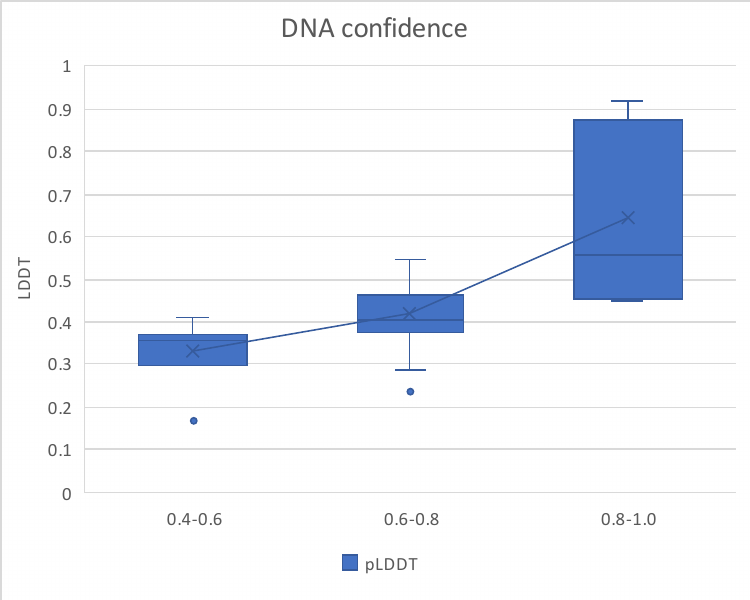}
        \caption{Relation between pLDDT and LDDT for DNA targets.}
        \label{fig:confidence_dna}
    \end{subfigure}
    \caption{Model confidence scores of HelixFold3.}
    \label{fig:confidence}
\end{figure}
Confidence scores from structure prediction models are essential for assessing the accuracy of their predictions. HelixFold3 employs several confidence metrics, including pLDDT, pAE, and pTM, to evaluate its predictions. We performed a correlation analysis between these confidence scores and the actual accuracy of predicted structures using data from protein complexes. HelixFold3 generated confidence scores for datasets, including small molecule ligand-protein interactions from PoseBusters, as well as protein-protein complexes, RNA molecules, and DNA molecules collected from the PDB. Across all these datasets, we observed a strong correlation between the confidence scores and the structural accuracy (Figure~\ref{fig:confidence}), indicating the reliability of these metrics in evaluating prediction quality.

\section{Conclusion}
Our team is rigorously developing HelixFold3 to replicate the capabilities of AlphaFold3. We reported our current progress, which shows that HelixFold3's accuracy on conventional ligands, nucleic acids, and proteins is approaching that of AlphaFold3. The inference code and current version of model parameters of HelixFold3 are open-sourced on GitHub to facilitate its use by researchers. We will continue to refine the model and will provide updates on HelixFold3’s performance with larger and more diverse datasets. We welcome you to stay updated on our progress. We invite you to follow our progress. For inquiries regarding HelixFold3 or potential commercial and research collaborations with the PaddleHelix team, please contact us at \texttt{\href{mailto:baidubio_cooperate@baidu.com}{baidubio\_cooperate@baidu.com}}.

\section*{Acknowledgement}
We sincerely acknowledge the invaluable support of computing resources from Tecorigin. Their contributions have been critical to the successful development of HelixFold3, advancing our efforts to further progress in life sciences.

% \clearpage
% \begin{appendices}
% \renewcommand{\thesection}{\arabic{section}}
% \renewcommand{\thetable}{\arabic{table}}
% \renewcommand{\thefigure}{\arabic{figure}}

% \section{Training}

% \section{Evaluation}

% \end{appendices}

\clearpage

%Bibliography
\bibliographystyle{unsrt}  
\bibliography{references}

\begin{thebibliography}{10}

\bibitem{jumper2021highly}
John Jumper, Richard Evans, Alexander Pritzel, Tim Green, Michael Figurnov, Olaf Ronneberger, Kathryn Tunyasuvunakool, Russ Bates, Augustin {\v{Z}}{\'\i}dek, Anna Potapenko, et~al.
\newblock Highly accurate protein structure prediction with alphafold.
\newblock {\em nature}, 596(7873):583--589, 2021.

\bibitem{evans2021protein}
Richard Evans, Michael O’Neill, Alexander Pritzel, Natasha Antropova, Andrew Senior, Tim Green, Augustin {\v{Z}}{\'\i}dek, Russ Bates, Sam Blackwell, Jason Yim, et~al.
\newblock Protein complex prediction with alphafold-multimer.
\newblock {\em biorxiv}, pages 2021--10, 2021.

\bibitem{abramson2024accurate}
Josh Abramson, Jonas Adler, Jack Dunger, Richard Evans, Tim Green, Alexander Pritzel, Olaf Ronneberger, Lindsay Willmore, Andrew~J Ballard, Joshua Bambrick, et~al.
\newblock Accurate structure prediction of biomolecular interactions with alphafold 3.
\newblock {\em Nature}, pages 1--3, 2024.

\bibitem{wang2022helixfold}
Guoxia Wang, Xiaomin Fang, Zhihua Wu, Yiqun Liu, Yang Xue, Yingfei Xiang, Dianhai Yu, Fan Wang, and Yanjun Ma.
\newblock Helixfold: An efficient implementation of alphafold2 using paddlepaddle.
\newblock {\em arXiv preprint arXiv:2207.05477}, 2022.

\bibitem{fang2023method}
Xiaomin Fang, Fan Wang, Lihang Liu, Jingzhou He, Dayong Lin, Yingfei Xiang, Kunrui Zhu, Xiaonan Zhang, Hua Wu, Hui Li, et~al.
\newblock A method for multiple-sequence-alignment-free protein structure prediction using a protein language model.
\newblock {\em Nature Machine Intelligence}, 5(10):1087--1096, 2023.

\bibitem{fang2024helixfold}
Xiaomin Fang, Jie Gao, Jing Hu, Lihang Liu, Yang Xue, Xiaonan Zhang, and Kunrui Zhu.
\newblock Helixfold-multimer: Elevating protein complex structure prediction to new heights.
\newblock {\em arXiv preprint arXiv:2404.10260}, 2024.

\bibitem{liu2023pre}
Lihang Liu, Donglong He, Xianbin Ye, Shanzhuo Zhang, Xiaonan Zhang, Jingbo Zhou, Jun Li, Hua Chai, Fan Wang, Jingzhou He, et~al.
\newblock Pre-training on large-scale generated docking conformations with helixdock to unlock the potential of protein-ligand structure prediction models.
\newblock {\em arXiv preprint arXiv:2310.13913}, 2023.

\bibitem{berman2000proteindatabank}
Helen~M Berman, John Westbrook, Zukang Feng, Gary Gilliland, Talapady~N Bhat, Helge Weissig, Ilya~N Shindyalov, and Philip~E Bourne.
\newblock The protein data bank.
\newblock {\em Nucleic acids research}, 28(1):235--242, 2000.

\bibitem{buttenschoen2024posebusters}
Martin Buttenschoen, Garrett~M Morris, and Charlotte~M Deane.
\newblock Posebusters: Ai-based docking methods fail to generate physically valid poses or generalise to novel sequences.
\newblock {\em Chemical Science}, 15(9):3130--3139, 2024.

\bibitem{kryshtafovych2023critical}
Andriy Kryshtafovych, Torsten Schwede, Maya Topf, Krzysztof Fidelis, and John Moult.
\newblock Critical assessment of methods of protein structure prediction (casp)—round xv.
\newblock {\em Proteins: Structure, Function, and Bioinformatics}, 91(12):1539--1549, 2023.

\bibitem{James2014sabdab}
James Dunbar, Konrad Krawczyk, Jinwoo Leem, Terry Baker, Angelika Fuchs, Guy Georges, Jiye Shi, and Charlotte M~Deane.
\newblock Sabdab: the structural antibody database.
\newblock {\em Nucleic acids research}, 42(D1):D1140--D1146, 2014.

\bibitem{chen2023rna}
Ke~Chen, Yaoqi Zhou, Sheng Wang, and Peng Xiong.
\newblock Rna tertiary structure modeling with briq potential in casp15.
\newblock {\em Proteins: Structure, Function, and Bioinformatics}, 91(12):1771--1778, 2023.

\bibitem{baek2024accurate}
Minkyung Baek, Ryan McHugh, Ivan Anishchenko, Hanlun Jiang, David Baker, and Frank DiMaio.
\newblock Accurate prediction of protein--nucleic acid complexes using rosettafoldna.
\newblock {\em Nature methods}, 21(1):117--121, 2024.

\end{thebibliography}

\end{document}